# Itinerant magnetism vis-à-vis structural phases in AlCuFeMn multi-principal-component medium entropy alloy

Palash Swarnakar[1], Partha Sarathi De[1], P D Babu[2], Amritendu Roy[1*]

School of Minerals, Metallurgical and Materials Engineering Indian Institute of Technology Bhubaneswar, Odisha-752050, India

*Corresponding author: amritendu@iitbbs.ac.in

**Abstract**

The exploration of multi-principal-component alloys (MPCAs) as potential functional materials in research is still in its early phase, with most studies centred on their potential application as structural materials. Magnetic materials possessing superior performance characteristics are essential for functional applications. Experimental observations and *ab-initio* density functional theory calculations were used to design and investigate an MPCA, AlCuFeMn, based on the medium-entropy effect. This study examines the microstructure evolution, phase formation, and soft magnetic behaviour of a cast and annealed AlCuFeMn MPCA. We conducted first-principles density-functional-theory (DFT) calculations to explain a selected multi-phase alloy's atomic, electronic, and magnetic structures at absolute zero temperature to understand the experimental findings better. We verified the predictions based on DFT by comparing them with the experimental observations. Despite the emphasis on equimolar compositions, the findings and conclusions of this study can enhance phase prediction and magnetic characteristics in non-equimolar alloys.

*Keywords*: Phase stability, Magnetism, Multi-principal-component alloy, DFT, EBSD, Rietveld refinement

## 1. Introduction

In contrast to traditional alloys, near-equiatomic multiple-component alloys exhibit an elevated configurational entropy and are commonly referred to as "high entropy alloys" (HEAs), complex, concentrated alloys (CCAs), or multi-principal-component alloys (MPCAs) [1,2]. Recently, MPCAs have been attracting attention as a promising alternative to conventional magnetic materials in various technological applications [3–13]. A few MPCAs, characterised by coexisting phases like FCC and BCC, showcase a combination of remarkable magnetic and mechanical properties [14–17]. Itinerant magnetism, characterised by the collective behaviour of conduction electrons contributing to magnetic ordering, plays a crucial role in understanding the magnetic properties of various materials, especially those exhibiting complex electronic structures such as in MPCAs. The interplay between itinerant magnetism and the inherent configurational disorder in MPCAs presents a unique platform for exploring novel magnetic phenomena.

It has been observed that MPCAs that include ferromagnetic (FM) elements such as Fe, Co, and Ni not only boost thermal stability but also improve magnetic properties [18–20]. According to Zhang et al. [18], the saturation magnetisation (Ms) and coercivity (Hc) of FeCoNi(AlSi)x alloys are affected by phase transitions. The magnetic properties of CoFeMnNix (x = Al, Cr, Ga, and Sn) MPCAs were analysed using *ab-initio* density functional theory (DFT) calculations by Zuo et al. [21]. The study revealed a significant increase in the magnetic moment of the alloy upon the addition of Mn to Al [21]. Researchers have discovered that several MPCAs containing aluminium form an ordered B2 phase, as revealed in the literature [3,21,22]. Nonetheless, the implications of B2 phase formation and the subsequent distribution of elements between the B2 and BCC phases on the soft magnetic properties of HEAs remain ambiguous. Some authors have reported an increase in saturation magnetisation in correlation with the increase in BCC and decrease in FCC, although the origin of this effect remains unidentified [18,23]. Additionally, it was discovered by Kao et al. [3] that the magnitude of Ms in AlxCoCrFeNi HEAs increased when transitioning from FCC to BCC crystal structure. The phase composition plays a crucial role in determining the magnetic properties of MPCAs. Nevertheless, our understanding of the influence of the phase constitution on the magnetic properties of HEAs is still limited.

MPCAs are expected to possess a solid solution composed of a single phase. In addition to the solid solutions (BCC, FCC, HCP) [1,2,24], there have been reports on intermetallic compounds, particularly Laves phases [25], and amorphous MPCAs [26], also known as bulk metallic glasses (BMGs). Considering this matter, it is imperative to consider additional sources of entropy, as they can be vital in maintaining phase stability in systems with multiple components. Recent experimental studies conducted by Roy et al. [27], Sharma et al. [28] and Dutta et al. [29] have extensively investigated the AlCuFeMn system, revealing its impressive high-temperature mechanical, oxidation, and corrosion characteristics. Roy et al. [27] reported the magnetic behaviour of the alloy AlCuFeMn; however, the role of phases in the magnetic property of the alloy was not explored. Fe and Mn, making up 25 at % each, are the magnetic elements in AlCuFeMn, which has a total composition of 50 at %. Additionally, at room temperature, AlCuFeMn MPCA demonstrates magnetic behaviour in both as-cast and annealed at 900 °C for 100 hours. Numerous factors, including processing and the material's thermal history, impact its magnetic properties [30]. Additionally, changes to both the quantity and crystal structure of the phases in the AlCuFeMn alloy occur when modifications are made to the processing conditions [27–29,31,32]. In this context, it is of paramount importance to comprehend the phase distribution and magnetic behaviour of AlCuFeMn MPCA.

Thus, there is an increased amount of research attention to this area. Equimolar AlCuFeMn, comprised of ferromagnetic Fe, also contains paramagnetic Al and Mn as well as diamagnetic Cu. The selection of Mn and Al here renders an interesting proposition since coexisting Al and Mn, albeit being non-magnetic, individually, have been shown to stabilise magnetic phase(s) [33]. Furthermore, copper (Cu) addition in equimolar FeCoNiAl alloy has been demonstrated to be effective in improving coercivity, albeit at the expense of saturation magnetisation [34]. In spite of such an interesting composition, during its short period of exploration, the magnetic characterisation of AlCuFeMn has not been studied in length. Roy *et al.* [27] only reported the temperature evolution of the magnetisation of the alloy. However, the role of magnetism in phase stabilisation *vis-à-vis* magnetic entropy was not studied, and the magnetic property of the alloy was not explored. In this context, it is of paramount importance to comprehend the phase distribution and magnetic behaviour of AlCuFeMn alloy.

## 2. Methodology

### 2.1 Experiments

Bulk AlCuFeMn alloy was melted and cast using commercial grade elements and ferroalloys in a vacuum induction melting-casting system under a controlled ambient of $10^{-4}$ bar. The cast ingot was annealed at 900 °C for 100 h. Structural characterisation was carried out at room temperature in a powder x-ray diffractometer (Bruker D8 Advance) in Bragg Brentano geometry using Cu-Kα radiation ($\lambda_{Cu-K\alpha} = 1.5418$ Å). FullProf Suite [35] was used for Rietveld refinement of the room-temperature pXRD data. Metallographic polishing of all samples was done using SiC papers and diamond paste, while the finishing was performed using 0.05 μm colloidal silica. The microstructure and chemical composition were investigated using field emission-scanning electron microscopy (FESEM) (Zeiss Merlin Compact) and energy dispersive spectroscopy (EDS) (Oxford Instruments). Magnetic measurements were carried out in a Physical Property Measurement System (PPMS) with a vibrating sample magnetometer (VSM) (Quantum Design). Temperature dependence of magnetisation (M – T) was studied over 10 to 300 K using both zero field cooled (ZFC) and Field cooled (FC) (field strength, 100 Oe, and 300 Oe) protocols. Magnetic field dependence of the magnetic moment was measured with a magnetic field, ± 7 Tesla, during the heating at temperatures 10, 100, 200, and 300 K. Magnetic domains were studied at room temperature using magnetic force microscopy (MFM) from Park Systems over a 5 μm×5 μm area.

### 2.2 Computation details

Vienna *Ab-initio* Simulation Package (VASP) [36,37] was used to perform first-principles calculations based on density functional theory (DFT) [38]. Perdew–Burke–Ernzerho exchange-correlation functional (GGA- PBE) was used [39]. A plane-wave cutoff energy of 550 eV, more than 1.5 times the maximum energy of the pure alloying components, was used throughout all calculations. The Brillouin zone samplings were done with the Monkhorst-Pack *k*-points [40], where the mesh grid density was fixed to $0.03 \times \frac{2\pi}{\text{Å}}$ for all the structures. The projector augmented wave (PAW) [41] pseudopotentials comprising 11 ($3d^{10}4s^1$) valence electrons for Cu, 8 ($3d^74s^1$) for Fe, 3 ($3s^23p^1$) for Al and 7 ($3d^54s^2$) for Mn were used. To address the exchange and correlation contributions of electrons to the Hamiltonian of the ion-electron system, the generalised gradient approximation (GGA) [39] was implemented. The self-consistent field (SCF) convergence criterion was set to $10^{-5}$ eV/atom, and the maximum force is 0.01 eV/atom. The first-order Methfessel–Paxton [42] approach with a broadening of 0.2 eV was used to relax all the structures. Spin-polarised calculations were performed using similar DFT parameters, considering the magnetic effects in the system at a temperature of 0 Kelvin.

A list of prospective phases in equimolar AlCuFeMn alloy vis-à-vis stability was studied in a previous work where consideration of magnetic entropy was not made [27,31]. Based on the experimental observation; it is likely that magnetisation has certain bearings in the phase stability of the alloy. In the present study, different magnetic states of the possible phases were considered, *viz,* ferromagnetic (FM), anti-ferromagnetic (AFM) and paramagnetic (PM). The PM state of the phases was treated using a disordered local magnetic moment (DLM) approximation [43]. Within DLM, the PM state of a phase in the alloy is modelled by splitting it into two with antiparallelly oriented spins with nonzero magnetic moments. The sum of the magnetic moments in the phase is preserved to be zero. While DLM has been traditionally used within the framework of coherent potential approximation (CPA), Alling et al. [44] successfully generalised the approach for supercell-based techniques. In the supercells of B2, L$2_1$, L$1_2$ and QH phases of AlCuFeMn alloy [31] magnetic moments of Fe and Mn atoms were arranged to simulate FM, AFM and PM states, respectively. In the ferromagnetic state, local magnetic moments attributed to Fe, and Mn atoms were arranged in parallel fashion, as shown in in Figure 1. In the second scenario, the initial alignment of local magnetic moments imitated the DLM state (paramagnetic) wherein the magnetic moments of Fe and Mn atoms display random orientation in relation to themselves (labelled as AlCu ($Fe_{\frac{1}{2}} \uparrow Fe_{\frac{1}{2}} \downarrow$) ($Mn_{\frac{1}{2}} \uparrow Mn_{\frac{1}{2}} \downarrow$) in Figure 1), resulting in zero net magnetisation. Based on the occupancy of the atoms, a group of arrangements are found to be possible for B2, L$2_1$ and QH phases (Figure 1). The presence of up and down arrows in the phase indicates the orientation of the local moments (Figure 1).

**Results and Discussion**

*2.1. Structural and microstructural characterisations*

Two regions were identified from the backscatter FESEM micrograph (Figure. 2 (a)) and the corresponding chemical compositional analysis (Figure 2 (b)): a Cu-rich bright dendritic region and a Fe-Mn-rich dark interdendritic region. The matrix shows a uniform distribution of Al. The chemical compositions have been tabulated in detail and are presented in Table 1.

According to the DFT study, it is predicted that at a temperature of 300 K, the equimolar AlCuFeMn MPCA would exhibit three phases: L$1_2$-1 ($Pm\bar{3}m$), L$2_1$ ($Fm\bar{3}m$), and B2-3 ($Pm\bar{3}m$), in order of increasing energy. The Rietveld method [45], as implemented in the FullProf suite [35], was used to refine the room temperature pXRD pattern of the bulk alloy to validate the theoretical predictions. The X-ray diffraction pattern (pXRD) and its corresponding fitting are displayed in Figure 2 (c). We documented the precise lattice parameters and additional crystallographic details in Table 2.

Phase fraction has been quantified through the analysis of FESEM micrographs, EBSD data, and the Rietveld refinement of pXRD. Based on the FE-SEM micrographs, it was determined that the Cu-rich region accounts for approximately ∼ 31.60 wt %, while the Fe-Mn-rich region makes up the remaining ∼ 68.40. According to the EBSD findings, B2-3 contains ∼ 16.85 wt

%, L2$_1$-3 ∼ 60.92 wt %, L1$_2$-1 ∼ 6.38 wt %, and the "zero solution" approximately ∼ 15.85 wt %. The Fe-Mn-rich L1$_2$-1 phase (Figure. 2 (d)) is mainly responsible for the occurrence of the "zero solution" in EBSD.

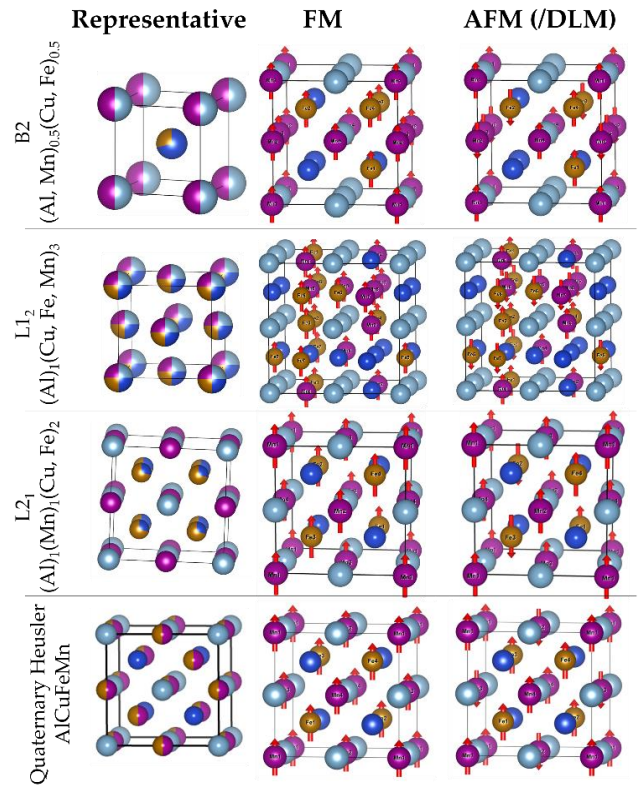

Figure 1: Schematic representation of B2-3, L1$_2$-1, L2$_1$-3, and Quaternary AlCuFeMn phases in FM and AFM states. State with net magnetic moment is equal to zero is defined as DLM state wherein half the magnetic ions.

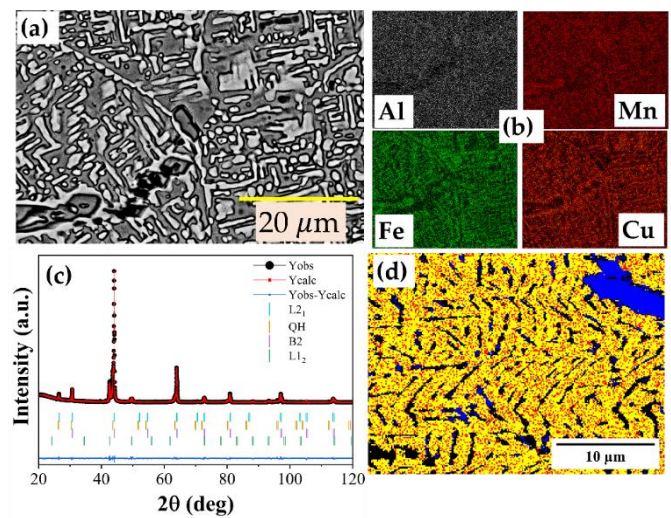

Figure 2: (a) SEM image in BSE detector, and the corresponding chemical compositions through EDS analysis (b), powder x-ray diffraction (pXRD) and the corresponding

Rietveld refinement with goodness-of-fitting fitting ($\chi^2$) ~ 5.52 and Rexp ~ 4.52, (d) EBSD micrograph in AlCuFeMn MPCA.

Table 1: Elemental distribution obtained from EDS analysis on the experimentally observable phases in the AlCuFeMn MPCA.

| Regions | Al (at. %) | Cu (at %) | Fe (at %) | Mn (at %) |
|---|---|---|---|---|
| Cu-rich | 22.62 ± 0.67 | 47.02 ± 5.13 | 14.59 ± 2.96 | 15.76 ± 1.75 |
| Fe-Mn-rich | 25.00 ± 0.49 | 12.90 ± 2.54 | 36.06 ± 2.12 | 26.00 ± 0.86 |
| Overall | 23.42 ± 0.18 | 26.03 ± 0.17 | 27.72 ± 0.06 | 22.84 ± 0.21 |

## *2.2. Magnetic Measurements*

Temperature-dependent magnetisation during cooling-heating cycles in the temperature range of 10-300 K is plotted in Figure 2(a). Both field-cooling and zero-field-cooling yielded a gradual increase in the magnetisation over the temperature domain, 300-100K, with the plots merging with one another. At a temperature close to ~100K, there is a sudden jump in magnetisation, typical of a paramagnetic to ferromagnetic transition. Ferromagnetic Curie temperature ($T_C$), estimated from the inflection point in the first-order temperature derivative of the *M(T)* curve as shown in the inset of Figure 2 (a) (top inset) is ~ 75 K. Below the $T_C$, irreversibility of the magnetic susceptibility is observed through the bifurcation of the FC and ZFC plots. Both the FC and ZFC plots reach a maximum at around 44 K ($T_m$) beyond which magnetisation, ZFC in particular, decreases sharply. Bifurcation of FC and ZFC plots and a sharp reduction in magnetisation (demonstrated by a second inflection point in dM/dT plot at ~25 K) could be variously attributed to factors such as (i) onset of another magnetic transition such as FM-AFM transition [46] (ii) onset of a magnetically frustrated spin state where spin fluctuations are slowed down upon cooling and consequently spins freeze in random directions in the absence of an external magnetic field, (iii) superparamagnetism, (iv) magnetocrystalline anisotropy or (v) a coupled phenomenon induced by the above [47,48].

Table 2: Possible phases observed in the room temperature pXRD analysis and the Rietveld refinement fitted lattice parameters (with DFT computed values) in AlCuFeMn MPCA.

| Phase | Space Group | Exp. $a = b = c$ (DFT) (Å) | Phase fraction (wt %) | Atomic coordinates | Occupancy |
|---|---|---|---|---|---|
| B2-3 | $Pm\bar{3}m$ (221) | 2.91 (2.91) | 26 | Al 1a $(0,0,0)$<br>Mn 1a $(0,0,0)$<br>Cu 1b $\left(\frac{1}{2},\frac{1}{2},\frac{1}{2}\right)$<br>Fe 1b $\left(\frac{1}{2},\frac{1}{2},\frac{1}{2}\right)$ | 0.50<br>0.50<br>0.70<br>0.30 |
| L1$_2$-1 | $Pm\bar{3}m$ (221) | 3.67 (3.69) | 9 | Al 1a $(0,0,0)$<br>Cu 3c $\left(0,\frac{1}{2},\frac{1}{2}\right)$<br>Fe 3c $\left(0,\frac{1}{2},\frac{1}{2}\right)$<br>Mn 3c $\left(0,\frac{1}{2},\frac{1}{2}\right)$ | 1.00<br>0.20<br>0.40<br>0.40 |
| QH | $F\bar{4}3m$ (216) | 5.82 (5.84) | ~1 | Al 4a $(0,0,0)$<br>Fe/Mn 4b $\left(\frac{1}{2},\frac{1}{2},\frac{1}{2}\right)$<br>Cu 4c $\left(\frac{1}{4},\frac{1}{4},\frac{1}{4}\right)$<br>Fe/Mn 4d $\left(\frac{3}{4},\frac{3}{4},\frac{3}{4}\right)$ | 1.00<br>0.50/0.50<br>1.00<br>0.50/0.50 |
| L2$_1$-3 | $Fm\bar{3}m$ (225) | 5.87 (5.81) | 64 | Al 4a $(0,0,0)$<br>Mn 4b $\left(\frac{1}{2},\frac{1}{2},\frac{1}{2}\right)$<br>Cu/Fe 8c $\left(\frac{1}{4},\frac{1}{4},\frac{1}{4}\right)$<br>Cu/Fe 8c $\left(\frac{3}{4},\frac{3}{4},\frac{3}{4}\right)$ | 1.00<br>1.00<br>0.30/0.70<br>0.30/0.70 |

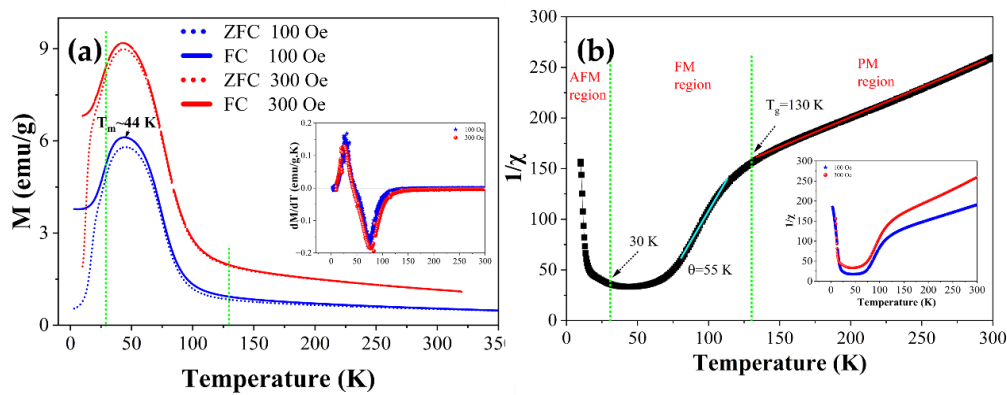

Figure 3: (a) FC-ZFC curve with d$M(T)$/dT as a function of temperature, and (b) $1/\chi$ as a function of temperature exhibiting deviation from Curie-Weiss law, presence of Griffith phase and thereby separation of regions based on spin orientation.

The sharp reduction in the susceptibility, $\chi(T)$, above $T_m$ could be described by the Curie-Weiss law based on the local-moment Heisenberg model:

$$\chi(T) = \frac{C}{T - \theta} \quad (1)$$

The inverse susceptibility plotted as a function of temperature in the bottom inset of Figure 2 displays Curie-Weiss behaviour at a temperature higher than $T_C$, indicating spin interactions and a calculated Curie-Weiss temperature, $\theta \sim 55$ K, in contrast with $T_C \sim 75$ K estimated from the temperature derivative of $M(T)$ curve. In the case of local Heisenberg spins with a g-factor of 2 and spin angular momentum quantum number S expressed in units of h/2π, the Curie constant can be represented in Gaussian CGS units as $C = \frac{N_A p_{eff}^2 \mu_B^2}{3k_B}$, where $N_A$ denotes Avogadro's number, $\mu_B$ signifies the Bohr magneton, $p_{eff}$ represents the effective moment per formula unit (f.u), and $k_B$ stands for Boltzmann's constant. The magnitude of $p_{eff} \approx \sqrt{8C}$, as determined from the Curie-Weiss fits, is approximately equal to 1.83 $\mu_B$. The departure of the $\frac{1}{\chi}$ vs. $T$ plot from the Curie–Weiss behaviour around ~ 130 K suggests the formation of ferromagnetic clusters in the paramagnetic region just above θ [49]. This deviation exemplifies the Griffiths phase that occupies the transitional region between the paramagnetic phase and the ferromagnetic phases [49]. Earlier investigations have demonstrated the differentiation of AFM-FM-PM regions with increasing temperature, as a result of the presence of the Griffith phase in several other systems [50,51].

While FM-AFM transition below ~20 K is a possibility, the observed low-temperature behaviour of the $M(T)$ may also be explained in terms of competing exchange interactions, potentially leading to magnetically frustrated behaviour. Alternatively, the presence of weakly interacting spin clusters freezing below ~20 K could also explain this phenomenon. Consequently, conducting magnetisation measurements on either side of the ZFC maximum may prove valuable in determining the magnetic ground state.

Figure 3 (a) displays magnetic isotherms at various temperatures. The graph illustrates that as the temperature decreases: (i) the magnetisation value rises, (ii) the magnetisation does not reach saturation even at a field of 60 kOe, and (iii) a nonlinear $M - H$ behaviour is observed even at 300 K, suggesting the existence of magnetic clusters well beyond the transition temperature as shown by the magnetic force microscopy (MFM) images shown in Figure 3(b) and (c) taken from a 5 μm×5 μm area of the sample. Figures 3(b) and (c) illustrate topographic and phase images where the magnetic region in the sample appeared to be of irregular shape, and there is a noticeable difference in magnetic contrast between the two types of regions. One can also see a large number of nano inclusions. These nano-phase particles can profoundly impact the magnetic structure since they can pin the motion of the domain walls.

As shown in Figure. 3(a), the $M(H)$ curves at $T \geq 100$ K exhibit negligible remanence and coercivity while displaying non-linear behaviour that implies the existence of magnetic clusters, which was subsequently observed in room-temperature MFM images. To estimate the clusters' momentum, the $M - H$ data was fitted to a modified Langevin function with an added linear term to accommodate the high field susceptibility:

$$M(H,T) = M_S L\left(\frac{\mu_p H}{k_B T}\right) + \chi_{hf} H$$
$$and\ L\left(\frac{\mu_p H}{k_B T}\right) = coth\left(\frac{\mu_p H}{k_B T}\right) - \frac{1}{\left(\frac{\mu_p H}{k_B T}\right)} \quad (2)$$

The term '$M_S$' represents the saturation magnetisation while $\mu_p$ represents the magnetic moment of the cluster in Bohr magnetons. The solid lines observed intersecting the data points in 3(d) indicate fitting to the Langevin function. The figure demonstrates excellent fits, indicating the presence of magnetic clusters above $T_C$. The magnetic moment of the cluster was determined to be approximately $2.72 \times 10^2$ $\mu_B$ implying the potential existence of a superparamagnetic (SPM) state. Nevertheless, the magnetic moment of the cluster is small in comparison to that of AlNiCo, which is estimated to be around $10^4$ [20]. The presence of the SPM state can be verified by specific characteristics: (a) magnetic isotherms should exhibit anhysteritic behaviour, where $M_R = 0$ and $H_c = 0$, (b) the magnetisation curve of an isotropic sample should exhibit temperature dependence to the extent that the curves obtained at different temperatures coincide when plotted against H/T above the temperature $T_p$. This phenomenon is evident in the scenario of an ideal non-interacting SPM behaviour. Figure 3(e) depicts the normalised magnetisation as a function of $H/T$ for temperatures ranging from 100 K to 300 K. It can be observed that the isotherms overlap in the low field range, but significant deviations are observed at higher fields. Additionally, it is worth noting from Figure 2 that the magnetisation curve of the $FC$ exhibits an upward turn beyond $T_m$ as the temperature is further reduced which is a signature of the nature of a single-domain paramagnet [52].

It is further noted that at temperature 10 K, the magnetic coercive field ($H_C$) measures approximately 360 Oe, which decreases to around 11 Oe at 100 K. Both $H_C$ and remanence ($M_r$) disappear at temperatures exceeding 100 K (refer to the inset of Figure 3(a)). It was also observed in Figure 2 that when a higher magnetic field was applied, the peak in the field-cooled (FC) curve at low temperature was suppressed (with respect to the ZFC peak). The application of higher magnetic fields reveals a substantial boost in magnetisation values. Consequently, by examining the magnetic isotherms captured at different temperatures, we calculated the spontaneous magnetisation ($M_{SP}$) using the following expression [20]:

$$M(H,T) = M_{SP} + A\sqrt{H} + \chi_{hf} H \quad (3)$$

The term $\sqrt{H}$ accounts for the suppression of spin waves by the field H, while $A$ remains constant. In Figure. 4 (a), the data for $M(T, H)$ is plotted alongside the least-squares fits. At 10 K, the values for $M_{SP}$ and the high-field susceptibility ($\chi_{hf}$) are determined to be 6.75 emu/g and $-2.32 \times 10^{-4}$ emu/g Oe, respectively. The spontaneous magnetic moment ($p_s$), represented as $\frac{M_{SP}\left(\frac{emu}{g}\right) \times MW}{5585}$, where $MW$ is the molecular weight, has been estimated to be 0.243 $\mu_B$ at a temperature, 10 K.

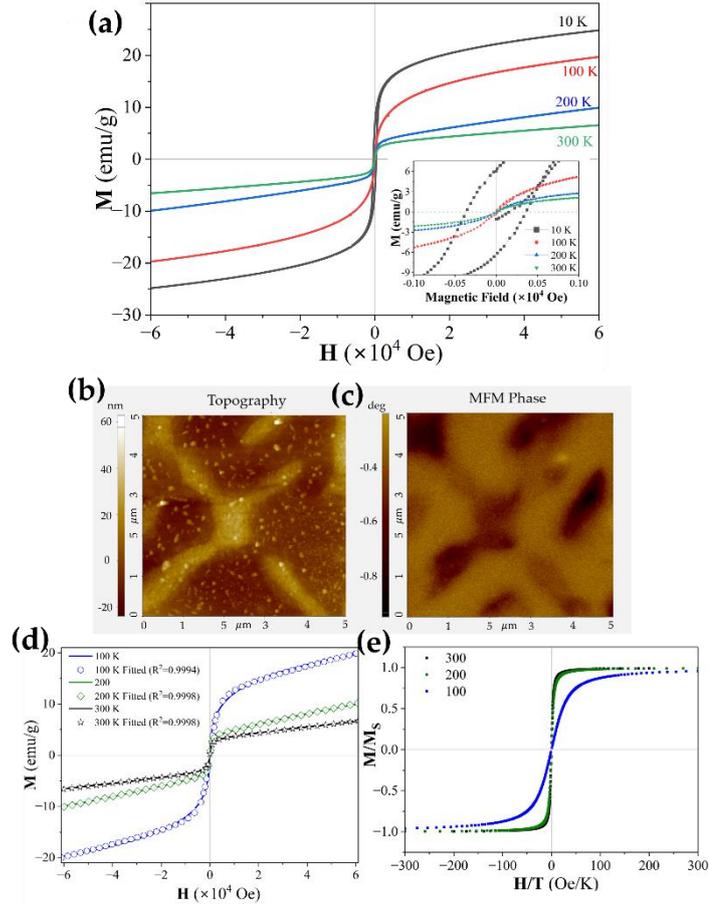

Figure 4: (a) Magnetic-hysteresis loops at temperatures 10, 100, 200, and 300 K in AlCuFeMn MPCA. A zoomed portion of the hysteresis loop (bottom part on the right side), (b) MFM images corresponding to topographic (c) and magnetic mapping (d) LAS fit to M vs. H data fitted to Eq. (2), and (e) normalised magnetisation vs H/T to verify the scaling behaviour.

It is worth noting that the value of $p_{eff}$ is comparatively greater than that of $p_s$. The magnetic carrier ($p_c$) concentration was determined using: $p_{eff}^2 = p_c(p_c + 2)$. Magnetic carrier concentration in an ordered state ($p_s$) was obtained through analysis of the low-temperature magnetic isotherm [20]. Further, Rhodes-Wohlfarth ratio ($RWR$) was determined by $\frac{p_c}{p_s}$, providing a measure of the change in magnetic carriers as temperature decreases. In the case of a localised system, if the moment remains relatively constant, the $RWR$ will approach unity. Conversely, the $RWR$ is significant for itinerant or delocalised systems [20]. The $RWR$ obtained for the AlCuFeMn is approximately 4.4, which is similar to a typical weak itinerant ferromagnet like ZrZn$_2$ (5.5) and significantly less than that of AlNiCo (14) [20]. This suggests the presence of a weak-itinerant character of ferromagnetism in the present alloy.

Further, by applying the law of approach to saturation (LAS), we determined the magnetic anisotropy ($K_1$). The empirical formula attributed to LAS is [20]

$$M = M_s \left[1 - \frac{a}{H} - \frac{b}{H^2}\right] + \chi H \tag{4}$$

The susceptibility ($\chi$) term appears to be unaffected by the metallurgical history of the material and is believed to originate from the variation in the intrinsic domain magnetisation with the applied magnetic field. The coefficients $a$ and $b$ are strongly influenced by the metallurgical history of the material. The value of $a$ exhibits a direct relationship with the density of dislocations influencing the extent of plastic deformation, whereas $b$ is primarily correlated with the crystalline properties of the material. In cubic crystal structures, the value of $b$ can be determined using the equation $b = \frac{8K_1^2}{105M_s^2}$, where $K_1$ represents the magnetic anisotropy constant and $M_s$ is the saturation magnetisation. The magnetisation isotherm data at 10 K is fitted to Eq. (5), allowing the determination of the magnetic anisotropy value. The coefficient, $b$ was estimated to be $-5.74 \times 10^6$ ergs/cm$^3$, which is in agreement with the observed high $H_C$ value and the divergence of $ZFC$ and $FC$ magnetization curves. The $K_1$ value for AlCuFeMn was determined to be around 10$^5$, which falls within the same order of magnitude as AlNiCo [20], AlNiCoCu [53], and AlNiCoFe [53].

*Takashi Model for WIFM*
In order to explain the WIFM, Takahashi took into account both thermal fluctuations and zero-point fluctuations (also known as quantum fluctuations). In accordance with Landau's theory, the expression for the free energy F(M) of an FM system is as follows:

$$F(M) = F(0) + \frac{1}{2}a_2(0)M^2 + \frac{1}{4}a_4(0)M^4 - MH \tag{5}$$

where $a_2(0) = 1/\chi(0)(2\mu_B)^2$ and $a_4(0) = F_{10}/N_0^3(2\mu_B)^4$, $F_{10}$ is a mode-mode coupling term. The role of the quartic M$^4$ term in equation (3) is crucial when it comes to the phase transition [54]. The minimisation of the free energy with respect to the magnetisation $[\partial F(M,T)/\partial M = 0]$ yields the relationship between $M$ and $H$ as:

$$H = \frac{F_{10}}{N_0^3(2\mu_B)^4}(-M_{sp}^2 + M^2)M \tag{6}$$

The expression for the mode-mode coupling term ($F_{10}$) can be represented in terms of the spin fluctuation parameters $T_0$, and $T_A$ in the following manner:

$$F_{10} = \frac{4T_A^2}{15T_0} \quad (7)$$

where $T_0$ corresponds to the distribution of dynamical spin fluctuations in the energy space while $T_A$ quantifies the breadth of the spin fluctuations spectrum distribution in wave space. The calculation of $F_{10}$ can be performed by analysing the slope of the Arrott plot ($M^2$ vs $H/M$) at low temperatures, as depicted in Figure 6.

$$F_{10} = \frac{N_A^3(2\mu_B)^4}{k_B \zeta} \quad (8)$$

The symbols $N_A$ and $k_B$ represent Avogadro's number and the Boltzmann constant, respectively, while $\zeta$ indicates the slope of the Arrott plot curve at low temperatures. The expression for the spontaneous magnetisation in the magnetic ground state is given in terms of $T_0$, $T_A$, and $T_C$ as

$$M_{sp}^2(0) = \frac{20T_0}{T_A} C_{4/3} \left(\frac{T_C}{T_0}\right)^{4/3} \quad (9)$$

The parameters $T_0$ and $T_A$ can be expressed as a function of $T_C$, $M_{sp}(0)$, and $F_{10}$ by employing equations (4)-(6), as shown.

$$\left(\frac{T_C}{T_0}\right)^{5/6} = \frac{M_{sp}^2(0)\sqrt{30C_Z}}{40C_{4/3}} \left(\frac{F_{10}}{T_C}\right)^{1/2} \quad (10)$$

$$\left(\frac{T_C}{T_A}\right)^{5/3} = \frac{M_{sp}^2(0)}{20C_{4/3}} \left(\frac{2T_C}{15C_Z F_{10}}\right)^{1/3} \quad (11)$$

Let $C_{4/3}(C_{4/3} = 1.006089)$ and $C_Z(C_Z = 0.5)$ be denoted as constants. Takahashi's theory holds great importance as it allows for the calculation of the $F_{10}$, $T_0$, and $T_A$ parameters exclusively using experimental magnetic data at low temperatures [48]. The calculation of spin fluctuations parameters is carried out using equation (7), and the results are presented in Table 1. These findings align well with the commonly observed WIFMs [55].

In our earlier calculation, we have computed Msp which is 6.75 emu/g which is 0.243 $\mu_B$/fu, and the slope ($\zeta$) from the M² vs H/M of 10 K data is 0.1389 (emu/g)³/Oe or 0.0648 $\mu_B$/T. The computed values for F01 found to be 0.288×10⁵ K, T0 is 1.04×10³ K, and TA was 1.06×10⁴ K. With these values the $\frac{T_c}{T_0} \approx 0.0719$ which is $\ll 1$ which is an indicator of WIFM.

*Magnetic entropy calculations*

Generally, the Arrott plot demonstrates a linear pattern in the high-field region, with the $M^2$ intercept providing the Magnetic Spontaneous Polarization ($M_{SP}$) at temperatures below the Curie temperature ($T < T_C$). When the value of $T$ reaches $T_C$, the straight line passes directly through the origin. On the other hand, when considering spin fluctuations, the $M^4$ vs $H/M$ relationship would display a line passing through the origin at $T_C$ [55]. An examination of the $M^4$ vs $H/M$ plots revealed a $T_C$ value that is consistent with the $M(T)$ data in Figure 2. The observations suggest that the temperature-induced spin fluctuations are crucial in comprehending the phase transition.

Intrinsic magnetic interaction vis-à-vis magnetocaloric characteristic of the alloy can be analysed by studying the change in magnetic entropy using Maxwell's relation:

$$\left[\frac{\partial S(T,H)}{\partial H}\right]_T = \left[\frac{\partial M(T,H)}{\partial T}\right]_H \quad (12)$$

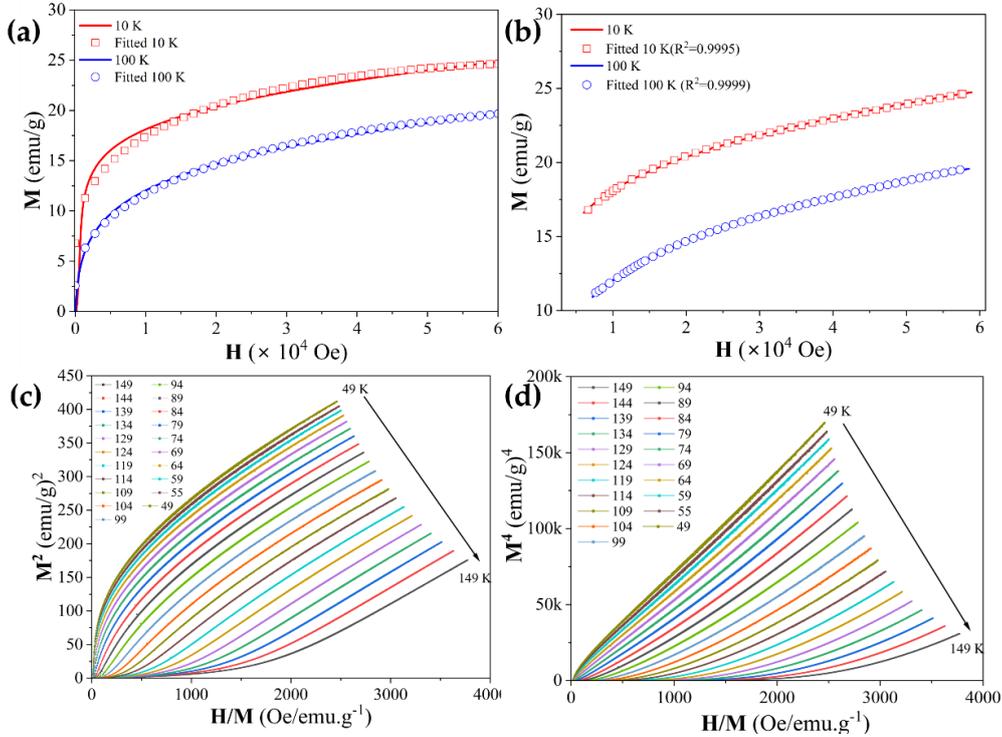

Figure 5: M vs. H data at temperatures 10 and 100 K, the line through the points are fits to Eq. (5) (a), Eq. (6) (b), (c) $M^2$ vs. $H/M$ as a function of temperature $50 < T < 150$, (d) $M^4$ vs. $H/M$ as a function of temperature $50 < T < 150$.

or, $$\Delta S_M(T,H) = S_M(T,H) - S_M(T,0) = \int_0^{H_{max}} \left[\frac{\partial M(T,H)}{\partial T}\right]_H dH \quad (13)$$

Using the magnetic isotherms measured at a discrete small magnetic field, as shown in Figure 5(a), $\Delta S_M(T,H)$ was calculated using [54]:

$$\Delta S_M(T_i, H) = \frac{\int_0^H M(T_i, H)dH - \int_0^H M(T_{i+1}, H)dH}{T_i - T_{i+1}} \tag{14}$$

From the $M - H$ data over the temperature range, 50 to 150 K (with a 5 K step), as shown in Figure. 5 (a), the magnetic entropy change ($-\Delta S_{mag}$) was calculated using eq. (7-8) and are plotted in Figure. 5(b) as a function of temperature at different applied field protocols, 0-5 T. The presence of a peak at the critical temperature ($T_C \approx 100$ K) in each curve indicates that the magnetic entropy change reaches its highest at $T_C$, regardless of the applied magnetic field. The maximum value of $-\Delta S_{mag}$, 0.69 J/kg.K, under an applied field of 0-1 T is comparable to those of FeCoNiCrAl alloy (0.674 J/kg.K under 0-0.82 T) [56]. $-\Delta S_{mag}^{max}$ exhibits a power law relationship with the magnetic field, $-\Delta S_{mag}^{max} = aH^{n'}$, where the coefficient $n'$ is a function of the magnetic state of the sample. Fitting the power law equation to the present data yields a value of $n' = 0.807$, as depicted in the Figure. 5 (c). Franco et al [57] stated that in the vicinity of $T_C$, $n'$ is found to be 0.75 for a second-order phase transition, while in the ferromagnetic (FM) region below $T_C$ it becomes 1, and in the paramagnetic (PM) region it equals 2. Consequently, it can be concluded that the magnetic transition in AlCuFeMn MPCA is a second-order phase transition (SOPT). Figure 5 further demonstrates that the $\Delta S_{mag}$ contribution is appreciable in AlCuFeMn alloy ($-\Delta S_{mag}$ value at 0-5 T is 2.58 J/kg.K, which is $\sim 0.25\ R$ (where $R$ is universal gas constant)). In light of the above experimental observations, it was deemed necessary to reassess the phase evolution in AlCuFeMn alloy [31] by considering the influence of magnetic entropy on the energetically favourable phases.

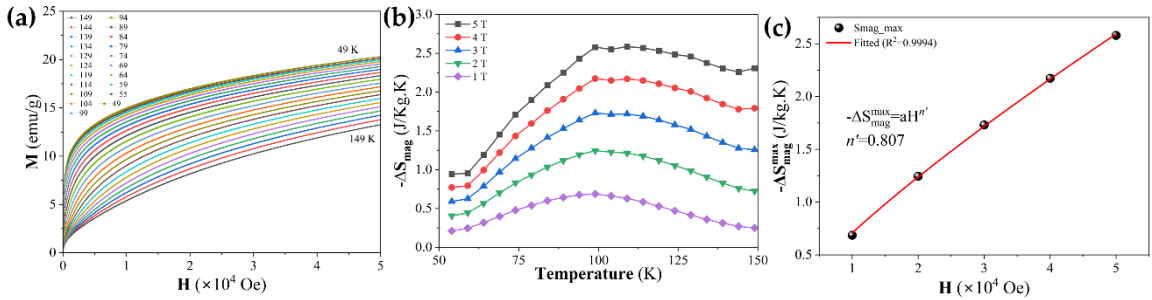

Figure 6: (a) Magnetic isotherm (M-H) data taken from 50 to 150 K with an interval of 5 K, (b) corresponding change in magnetic entropy ($-\Delta S_{mag}$) with temperature, and (c) non-linear fit of change in magnetic with a magnetic field.

## 2.3. First-principles studies

The nature of magnetism in multi-phase, complex alloys such as in AlCuFeMn [27–29] would depend on the degree of itinerancy of the constituent magnetic phases present in the alloy. While in highly itinerant systems, stable magnetic moments on magnetic sites do not survive, and long-range magnetism is due to the conduction electrons, in the local magnetic moment limit, stable magnetic moments are found at atomic sites. In the present work, considering a weak itinerancy (as observed above), we first assumed the local magnetic moment picture describing magnetism in the alloy. Thus, to understand the long-range magnetic order in the alloy and to evaluate its effect on the structural phase evolution, we computed and compared the Gibbs free energy of mixing of the competing phases determined within the density functional theory framework.

Based on the findings of our previous study [31], it was determined that the Cu-rich phase exhibited an ordered BCC structure, i.e., B2. The Fe-Mn-rich phase comprised two FCC structures, with an $L2_1$ matrix and nanostructured $L1_2$ (Figure. 3 (a-b)) particulates. In our previous study, Gibbs free energy versus temperature (G−T) plots (without consideration of magnetic entropy), did not demonstrate any transitions at the specified transition temperatures. It is probable that the exclusion of $S_{mag}$ (and corresponding $F_{mag}$) in the G − T has been responsible for this.

In high-entropy alloys, the Gibb's free energy of mixing can be calculated as

$$G_{mix} = H_{mix} - TS_{mix} \qquad (15)$$

The computation of the mixing enthalpy, $H_{mix}$, can be represented by the DFT static energy of a system at ($E^{sys}$) 0 K. The entropy term is the combination of configurational, vibrational, electronic, magnetic, and lattice mismatch entropy, i.e., $S_{mix} = S_{conf} + S_{vib} + S_{mag} + S_{lat}$.

Therefore, the Gibb's free energy of mixing can now be formulated as

$$G_{mix} = [E^{sys}] - T[S_{conf} + S_{vib} + S_{mag} + S_{lat}] \qquad (16)$$

Tables S1 and S2 in the supplementary material contain comprehensive details of the likely phases.

The magnetic entropy to the free energy can be [45] determined using:

$$F_{mag}(V,T) \approx -TS_{mag}(V) \qquad (17)$$

where magnetic entropy $S_{mag}$ is

$$S_{mag}(V) = k_B \sum_{i=1}^{N} c_i \ln(|M_i(V)| + 1) \qquad (18)$$

where $c_i$, and $M_i$ are the concentration, and average magnetic moment of the element $i$, $N$ is the total number of principal components.

Following the computation of DFT static energies ($E^{sys}$) at 0K, a selection process was undertaken to eliminate fewer probable phases, whereby the structure with the most negative $E^{sys}$ was positioned at the top of the list of possible phases. Subsequently, the dynamical stability of the probable phases was studied by performing lattice dynamics (phonon) calculations [46]. The Helmholtz free-energy $F(V,T)$ can be represented as a function of volume and temperature under the harmonic approximation:

$$F(V,T) = F_{vib}(V,T) + F_{el}(V,T) - T(S_{conf} + S_{mag} + S_{lat}) \qquad (19)$$

where $E^{sys}$ is the static total free energy at 0 K, $F_{el}(V,T)$ is the electronic free energy, $F_{vib}(V,T)$ is the vibrational free energy, $S_{conf}, S_{lat}$ and $S_{mag}$ are entropies concerning the configurational, lattice mismatch and magnetism of a system, respectively [31]. Under the quasi-harmonic approximation (QHA) [58], the Gibbs free energy of competing phases was computed as a function of temperature. The Gibbs free energy was determined by transforming the Helmholtz free energy [58].

$$G(p,T) = \min_{V}[F(V,T) + pV] \qquad (20)$$

In order to compare all potential magnetic/spin configurations, we considered the ferromagnetic (FM) state of magnetic atoms to include the magnetic contribution to phase stability. Table 2 displays the $E^{sys}$ energies of all possible configurations. phases characterised by lower static energies (i.e., more negative values) were identified as viable candidates for further examination. In addition, the temperature-dependent electronic and vibrational free energies of the configurations were analysed and compared, as shown in Figure. S3. The subsequent screening primarily involved studying the ferromagnetically stable structures, with a particular emphasis on their anti-ferromagnetic (AFM) and paramagnetic (or disorder local moment (DLM)) states, Table S3. The structures thus predicted from the configurational study were further compared based on their relative stability as a function of temperature, employing the Gibbs free energy formalism, as discussed in Eq. 8 and depicted in Figure. 2.

In order to assess the presence of anti-ferromagnetism in the L2$_1$, B2 and QH phases, as observed in Figure. 2 (b), we explored the potential AFM configurations in these two phases. The Supplementary File provides a comprehensive listing of the configurations and the computed energy per atom for all possible configurations at 0 K.

The considerable diminishment in energetic disparity between competing configurations (FM, DLM, AFM) across all phases (B2, L1$_2$, QH, L2$_1$) hinders any assessment of the phase stability in relation to potential magnetic configurations. However, experimental observations and DFT energies at 0 K indicate that Mn atoms have a strong tendency to align ferromagnetically with other Mn atoms, regardless of the arrangement of Fe atoms (whether ferromagnetic or antiferromagnetic) in the L2$_1$ structure. In the B2 structure, Mn atoms align in an antiparallel manner, and therefore, resulting in the formation of two separate magnetic regions in MFM images. The findings from the M-T studies and powder XRD indicate the imperative coexistence of three phases: paramagnet L2$_1$, B2, and ferromagnetic L1$_2$, at room temperature (300 K). Furthermore, the electronic configuration of QH phase, being $2 \times 2 \times 2$ of B2 renders its distinct identification highly challenging.

Table 3: Computed formation energies and configurational, lattice mismatch, and magnetic entropies for dynamically stable phases in AlCuFeMn MPCA.

| System | $E^{sys}$ (eV/atom) | $S_{conf}$ ($k_B$/atom) | $S_{lat}$ ($k_B$/atom) | $S_{mag}$ ($k_B$/atom) |
|---|---|---|---|---|
| L2$_1$ (DLM) | -6.2923 | 0.347 | 0.131 | 0.000 |
| L2$_1$ (FM) | -6.2948 | 0.347 | 0.131 | 1.042 |
| L2$_1$ (AFM) | -6.2946 | 0.347 | 0.131 | 1.043 |
| QH (AFM) | -6.2858 | 0.000 | 0.131 | 0.000 |
| QH (DLM) | -6.2858 | 0.000 | 0.131 | 0.000 |
| B2 (FM) | -6.2570 | 0.693 | 0.081 | 1.038 |
| QH (FM) | -6.2569 | 0.000 | 0.131 | 1.099 |
| L1$_2$ (FM) | -6.2703 | 0.824 | 0.131 | 1.244 |
| B2 (DLM) | -6.2759 | 0.693 | 0.081 | 0.016 |
| B2 (AFM) | -6.2765 | 0.693 | 0.081 | 0.061 |
| L1$_2$ (DLM) | -6.2314 | 0.824 | 0.131 | 0.337 |

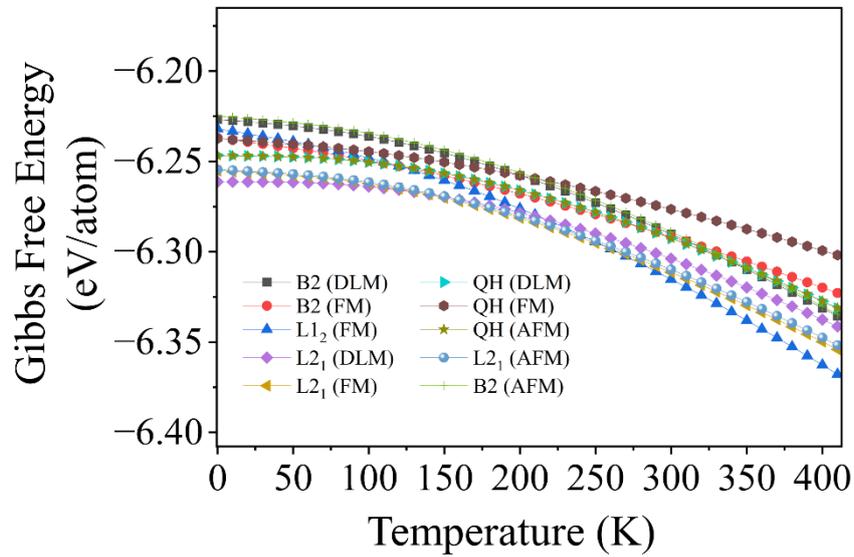

Figure 7: Change in Gibbs free energy of probable phases in AlCuFeMn alloy.

## 3. Conclusions

In summary, the experimental and ab-initio DFT studies conducted in this research explore the structural and magnetic properties of a new class of MPCA, AlCuFeMn. By comparing the Gibbs free energy of competing phases, we predicted three probable phases: B2 $(Al, Mn)_{0.5}(Cu, Fe)_{0.5}$, L2$_1$ $(Al)_1(Mn)_1(Cu, Fe)_2$, and L1$_2$ $(Al)_1(Cu, Fe, Mn)_3$. We verified the predicted phases through the room-temperature powder x-ray diffraction, FESEM microstructures, and electron backscatter diffraction. Analysis of the microstructure using FE-SEM, and elemental compositional mapping has unveiled two distinct regions, specifically Cu-rich and Fe-Mn-rich. The temperature-dependent M−T studies show a transition from strong to weak ferromagnetism within the temperature range of 10 K to 300 K. The magnetic force microscopy (MFM) studies at room temperature (300 K) verify the existence of two magnetically distinct domains exhibiting opposite spins. By combining FESEM (with elemental mapping) and MFM (with topography), we have confirmed that the Cu-rich region is down-spin magnetic, and the Fe-Mn-rich region is up-spin magnetic. According to the results obtained from EBSD and early TEM, it is observed that the Cu-rich region exhibits the B2 $(Al, Mn)_{0.5}(Cu, Fe)_{0.5}$ phase, whereas the Fe-Mn-rich region is a fused mixture of phases, namely L2$_1$ $(Al)_1(Mn)_1(Cu, Fe)_2$ and L1$_2$ $(Al)_1(Cu, Fe, Mn)_3$. The phase fractions are determined using pXRD, providing evidence that the Fe-Mn-rich region occupies larger domains. As determined from DFT studies and observed experimentally, the average magnetic moments exhibit corroborative agreement. Despite the different orientations of magnetic domains in Cu-rich and Fe-Mn-rich regions, the distinct segregation of magnetic elements in these phases is accountable for the weak ferromagnetism observed at room temperature (∼ 300 K). Furthermore, incorporating a more advanced and detailed characterisation approach, namely neutron diffraction, might aid in understanding the phase evolution and conducting a magnetic domain study in AlCuFeMn MPCA.

# Appendix A. Literature related to AlCuFeMn magnetic measurements.

Table A.1: Synthesis condition (or heat treatment), phases, saturation magnetisation ($M_s$), coercivity ($H_c$) of MPCAs. AC, MA, LENS represents as-cast, mechanical alloying, and laser engineered net shaping, respectively. All observations are at 300 K.

| | Synthesis | Phases | $M_s$ | $H_c$ | References |
|---|---|---|---|---|---|
| FeCoNi | AC | FCC | 163.4 | — | [16] |
| FeCoNi(AlCu)0.2 | AC | FCC | 134.3 | — | [16] |
| FeCoNi(AlCu)0.6 | AC | FCC | 90.7 | — | [16] |
| FeCoNi(AlCu)x (x = 0.7-1.2) | AC | FCC+BCC | 73-90 | — | [16] |
| FeCoNiAl0.25Mn0.25 | AC | FCC | 101 | 3.4 | [15] |
| FeCoNiAl0.5Mn0.5 | AC | FCC+BCC | 51.9 | 9.2 | [15] |
| FeCoNiAl0.75Mn0.75 | AC | FCC+BCC | 129.6 | 5.6 | [15] |
| FeCoNiAlMn | AC | BCC | 132.2 | 3.3 | [15] |
| Co0.5CrCuFeMnNi | MA | FCC+BCC | 21 | 63 | [49] |
| AlxCrCuFeNi2 (x = 0–1.5) | LENS | FCC+BCC | 0.8–21 | 10–74 | [50] |
| AlCoCrCuFeNi | Arc melting | FCC+BCC | 38.18 | 45 | [51] |
| AlCrFeMnNiTi | MA | FCC+BCC | 17.55 | 153.8 | [52] |
| FeCoNiCuMn | MA | FCC | 84 | 6 | [53] |
| FeNiCuMnCoCrx (x = 0.5-2.0) | MA | FCC+BCC | 25-61 | 8-75 | [53] |
| FeNiCuMnCrCox (x = 0.5-1.5) | MA | FCC+BCC | 21-40 | 19-63 | [53] |
| FeCoNiMn | AC | FCC | 18.14 | 1.5 | [22] |
| FeCoNiAlMn | AC | BCC+B2 | 147.86 | 7.9 | [22] |
| FeCoNiGaMn | AC | FCC+B2 | 80.43 | 11.5 | [22] |
| FeCoNiSnMn | AC | $L2_1$+BCC | 80.29 | 43.2 | [22] |
| FeCoNiAlCu | AC | FCC+BCC | 78.7 | 12.6 | [54] |
| FeCoNi1.25AlCu | AC | FCC+BCC | 75.6 | 4.8 | [54] |
| FeCoNi1.50AlCu | AC | FCC+BCC | 59.3 | 2.2 | [54] |
| FeCoNi1.75AlCu | AC | FCC | 54.4 | 2.1 | [54] |
| AlCuFeMn | AC+900ºC 100h | FCC + BCC | 7.08 | 6.92 | Present Study |

**Supplementary Material**

Please refer to the supplementary material for further details.

**Acknowledgements**

PS thanks the Ministry of Education, Government of India (GOI) for the fellowship. This work was supported by the UGC-DAE Consortium for Scientific Research through project number CSR CRS/2021-22/03/571.

**Data Availability**

The data that support the findings of this study are available from the corresponding author upon reasonable request.